\documentclass{amsart}
\usepackage{amssymb}

\usepackage{graphicx}

\input{tcilatex}
\input tcilatex

\begin{document}
\title[constant population size on average]{Wright-Fisher-like models with constant population size on average}
\author{Nicolas Grosjean and Thierry Huillet$^{*}$}
\address{Laboratoire de Physique Th\'{e}orique et Mod\'{e}lisation CNRS-UMR 8089 et
Universit\'{e} de Cergy-Pontoise, 2 Avenue Adolphe Chauvin, 95302,
Cergy-Pontoise, FRANCE\\
E-mail(s): Nicolas.Grosjean@u-cergy.fr and Thierry.Huillet@u-cergy.fr}
\maketitle

\begin{abstract}
We first recall some basic facts from the theory of discrete-time Markov
chains arising from two types neutral and non-neutral evolution models of
population genetics with constant size. We then define and analyse a version
of such models whose fluctuating total population size is conserved on
average only. In our model, the population of interest is seen as being
embedded in a frame process which is a critical Galton-Watson process. In
this context, we address problems such as extinction, fixation, size of the
population at fixation and survival probability to a bottleneck effect of
the environment.\newline

\textbf{Running title:} constant population size on average\newline


$^{*}$ corresponding author.
\end{abstract}

\section{Introduction}

Forward evolution of neutral large populations in genetics has a long
history, starting in the $1920$s; it is closely attached to the names of R.
A. Fisher and S. Wright; see (T. Nagylaki, $1999$) for historical
commentaries. The book of (W. Ewens, $2004$) is an excellent modern
presentation of the current mathematical theory. The starting point of such
neutral theories is embedded in the theory of discrete Markov chains whose
transition matrices are obtained from branching Galton-Watson processes
conditioned on keeping the total population size constant (as defined in
Karlin-McGregor, $1964$). Coalescent theory is the corresponding backward
problem, obtained while running the forward neutral evolution processes
backward-in-time. It was discovered independently by several researchers in
the $1980$s, but definitive formalization is commonly attributed to (J.
Kingman, $1982$). Major contributions to the development of coalescent
theory were made (among others) by P. Donnelly, R. Griffiths, R. Hudson, F.
Tajima and S. Tavar\'{e} (see the course of Tavar\'{e} in Saint-Flour $2004$
for a review and references therein). The neutral theory has been enriched
while including various drifts (or bias) describing say mutation,
recombination, selection effects superposing to the genetic drift...
Space-time scaling limits of such theories turn out to be very rich but in
this manuscript we shall stick to the discrete space-time setting. All such
recent developments and improvements concern chiefly the discrete neutral
case and their various scaling limits in continuous time and/or space. As
was shown for instance by (M\"{o}hle, $1994$ and $1999$), neutral forward
and backward theories learn much from one another by using a concept of
duality introduced by (T. Liggett, $1985$). There is therefore some evidence
that the concept of duality could help one understand the backward theory
even in non-neutral situations when various evolutionary forces are the
causes of deviation to neutrality (J. Crow and M. Kimura, $1970$); (T.
Maruyama, $1977$), (J. Gillepsie, $1991$) and (W. Ewens, $2004$), for a
discussion on various models of utmost interest in population genetics).

In this manuscript, after recalling some of the (forward and backward)
theory of the discrete space-time neutral genetic drift with two types (or
alleles), we briefly reconsider the case including various bias describing
deviation to neutrality. Various examples are discussed to fix the
background where we stress that the marginal populations of both types turn
out to be Markovian, even in the biased case. As emphasized earlier, all
this body of theory assumes an evolution process keeping constant over the
generations the total size of the population. This condition that to form
the next generation the offspring should preserve exactly the population
size to constant is a drastic one and the purpose of this work is to discuss
one way to understand a weaker form of conservation, namely conservation of
the total population size on average only. And see how the previous theory
is modified both in the neutral and the non-neutral cases with a bias
included. In our approach, a critical Galton-Watson process plays a key role.

Let us summarize our results: as in the constant population size context, we
first consider a two-types neutral population model in discrete-time. Type $%
1 $ population is seen as a subpopulation of a frame (or environment)
process which is modeled as a critical Galton-Watson process whose size is
constant over time, say $n$, but on average only. This process is already
present in the Karlin-McGregor way to handle a constant population size $n$
problem since, as $n$ goes to infinity, such models boil down to a critical
Galton-Watson process. The frame process exhibits very large growing
fluctuations and it has long-range positive correlations. The type $1$
population process also is a critical Galton-Watson process whose initial
condition is a subset of size $m$ of the $n$ founders generating the frame
process. We define the type $2$ population as the one whose founders is made
of the remaining subset of founders, therefore of size $n-m$. At all times
therefore, the type $1$ and $2$ populations, as embedded sub-processes
within the frame, sum up to the frame process itself. We then address the
following problems in this setup: what is the extinction probability of type 
$1$ population? And what is its fixation probability? For the former
problem, a first point of view is the classical one for critical branching
processes: extinction of type $1$ population is when it first hits state $0$%
. Such an extinction event occurs with probability $1$ but it takes a very
long time to do so. For the latter problem, we say that a type $1$ fixation
event occurs at some time if this time is the first at which type $2$
population goes extinct while type $1$ population remains alive. We compute
the probability distribution of the fixation time, together with the
probability that a fixation event takes place. We find that the fixation
time has finite mean (of order $n$ when both $m,n$ are large or when $m$ is
fixed and $n$ large), whereas its variance is infinite. The fixation
probability itself is found to be of order $m/n$ under the same assumptions.

Then we address the following problem: what is the size of type $1$
population at fixation whenever this event occurs? We find that on average
it is $m$, while its variance is infinite. Finally we discuss an alternative
and more symmetric way of defining type $1$ population extinction as the
fixation of the type $2$ population in the latter sense. With these
definitions at hand, one can compute the probability that a fixation event
of type $1$ precedes an extinction event and conversely.

We then observe that, under the neutral hypothesis, type $1$ and frame
processes, although not independent, are both Markov processes both
marginally and jointly. Inspired by similar ideas in the context of constant
population size models, we then introduce bias (as a deviation to
neutrality) into our model with fluctuating total population size. We show
that with this bias included, type $1$ and frame processes (still the same
critical mean-$n$ Galton-Watson process as in the neutral case) are still
jointly Markov but the type $1$ marginal process no longer is Markovian.

Finally, we briefly discuss the problem of defining an effective population
size in our model with variable population size and we investigate the
related question of computing the type $1$ survival probability to a
bottleneck effect of the frame process. In this context, the critical
homographic Galton-Watson model, which is invariant under iterated
composition, is shown to be of particular interest.

\section{Discrete-time neutral genetic drift and coalescent: a reminder}

In this Section, to fix the background and notations, we review some
well-known facts from the cited literature.

\subsection{Exchangeable neutral population models: Reproduction laws
examples.}

(The Cannings, $1974$ model). Consider a population with non-overlapping
generations $r\in \Bbb{Z}.$ Assume the population size is constant, say with 
$n$ individuals (or genes) over the generations. Assume the random
reproduction law at generation $0$ is $\mathbf{\nu }_{n}:=\left( \nu
_{1,n},...,\nu _{n,n}\right) ,$ satisfying: 
\begin{equation*}
\sum_{m=1}^{n}\nu _{m,n}=n.
\end{equation*}
Here, $\nu _{m,n}$ is the number of offspring of gene $m.$ We avoid the
trivial case: $\nu _{m,n}=1$, $m=1,...,n.$ One iterates the reproduction
over generations, while imposing the following additional assumptions:

- Exchangeability: $\left( \nu _{1,n},...,\nu _{n,n}\right) \overset{d}{=}%
\left( \nu _{\sigma \left( 1\right) ,n},...,\nu _{\sigma \left( n\right)
,n}\right) ,$ for all $n-$permutations $\sigma $ $\in \mathcal{S}_{n}.$

- time-homogeneity: reproduction laws are independent and identically
distributed (iid) at each generation $r\in \Bbb{Z}.$

This model therefore consists of a conservative conditioned branching
Galton-Watson process in $\left[ n\right] ^{\Bbb{Z}}$, where $\left[
n\right] :=\left\{ 0,1,...,n\right\} $ (see Karlin-McGregor, $1964$).\newline

Famous reproduction laws are:

\emph{Example 1} The $\theta -$multinomial Dirichlet family: $\mathbf{\nu }%
_{n}\overset{d}{\sim }$ Multin-Dirichlet$\left( n;\theta \right) $, where $%
\theta >0$ is some `disorder' parameter. With $\mathbf{k}_{n}:=\left(
k_{1},...,k_{n}\right) $, $\mathbf{\nu }_{n}$ admits the following joint
exchangeable distribution on the simplex $\left| \mathbf{k}_{n}\right|
:=\sum_{m=1}^{n}k_{m}=n$: 
\begin{equation*}
\mathbf{P}\left( \mathbf{\nu }_{n}=\mathbf{k}_{n}\right) =\frac{n!}{\left[
n\theta \right] _{n}}\prod_{m=1}^{n}\frac{\left[ \theta \right] _{k_{m}}}{%
k_{m}!},
\end{equation*}
where $\left[ \theta \right] _{k}=\theta \left( \theta +1\right) ...\left(
\theta +k-1\right) $ is the rising factorial of $\theta $. This distribution
can be obtained by conditioning $n$ independent mean $1$ P\`{o}lya
distributed random variables $\mathbf{\xi }_{n}=\left( \xi _{1},...,\xi
_{n}\right) $ on summing to $n$, that is to say: $\mathbf{\nu }_{n}\overset{d%
}{=}\left( \mathbf{\xi }_{n}:\left| \mathbf{\xi }_{n}\right| =n\right) ,$
where, with $\Bbb{N}_{0}=\left\{ 0,1,...\right\} $, 
\begin{equation*}
\mathbf{P}\left( \xi _{1}=k\right) =\frac{\left[ \theta \right] _{k}}{k!}%
\left( 1+\theta \right) ^{-k}\left( \theta /\left( 1+\theta \right) \right)
^{\theta }\text{, }k\in \Bbb{N}_{0}
\end{equation*}
or, equivalently in terms of its probability generating function (pgf), 
\begin{equation*}
\mathbf{E}\left( z^{\xi _{1}}\right) =:\phi \left( z\right) =\left( 1-\frac{1%
}{\theta }\left( z-1\right) \right) ^{-\theta }\text{, }z<z_{c}:=1+\theta .
\end{equation*}
P\`{o}lya (or negative binomial) distributed random variables are known to
be compound-Poisson (or infinitely divisible), meaning (Steutel and van
Harn, $2003$) 
\begin{equation*}
\phi \left( z\right) =e^{-\lambda \left( 1-\psi \left( z\right) \right) },
\end{equation*}
with $\lambda =\theta \log \left( 1+1/\theta \right) >0$ and with $\psi
\left( z\right) $ a pgf obeying $\psi \left( 0\right) =0$ and $\psi ^{\prime
}\left( 1\right) =1/\lambda $.

When $\theta =1$, $\xi _{1}$ is geometric and $\mathbf{\nu }_{n}$ is
uniformly distributed on the simplex $\mathbf{k}_{n}:\left| \mathbf{k}%
_{n}\right| =n$, with $\mathbf{P}\left( \mathbf{\nu }_{n}=\mathbf{k}%
_{n}\right) =1/\binom{2n-1}{n}.$

When $\theta \rightarrow \infty $, this distribution reduces to the
Wright-Fisher model for which $\mathbf{\nu }_{n}\overset{d}{\sim }$ Multin$%
\left( n;1/n,...,1/n\right) .$ Indeed, $\mathbf{\nu }_{n}$ admits the
following joint exchangeable multinomial distribution on the simplex $%
\mathbf{k}_{n}:\left| \mathbf{k}_{n}\right| =n$: 
\begin{equation*}
\mathbf{P}\left( \mathbf{\nu }_{n}=\mathbf{k}_{n}\right) =\frac{n!\cdot
n^{-n}}{\prod_{m=1}^{n}k_{m}!}.
\end{equation*}
This distribution can be obtained by conditioning $n$ independent mean $1$
Poisson distributed random variables $\mathbf{\xi }_{n}=\left( \xi
_{1},...,\xi _{n}\right) $ on summing to $n$. When $n$ is large, using
Stirling formula, it follows that $\mathbf{\nu }_{n}\overset{d}{\underset{%
n\rightarrow \infty }{\rightarrow }}\mathbf{\xi }_{\infty }$ with joint
finite-dimensional Poisson law: $\mathbf{P}\left( \mathbf{\xi }_{n}=\mathbf{k%
}_{n}\right) =\prod_{m=1}^{n}\frac{e^{-1}}{k_{m}!}=\frac{e^{-n}}{%
\prod_{m=1}^{n}k_{m}!}$ on $\Bbb{N}_{0}^{n}.$ Thanks to the product form of
all finite-dimensional laws of $\mathbf{\xi }_{\infty }$, we get an
asymptotic independence property of $\mathbf{\nu }_{n}$.

A slight extension of the P\`{o}lya model would be to consider a compound
P\`{o}lya model (and compound Poisson as well) for which, for some $z_{c}>1,$%
\begin{equation*}
\mathbf{E}\left( z^{\xi _{1}}\right) =:\phi \left( z\right) =\left( 1-\frac{1%
}{\theta h^{\prime }\left( 1\right) }\left( h\left( z\right) -1\right)
\right) ^{-\theta }\text{, }z<z_{c},
\end{equation*}
with $h\left( z\right) $ some pgf obeying $h\left( 0\right) =0$, $%
h^{^{\prime }}\left( 1\right) >1$ and $h^{^{\prime \prime }}\left( 1\right)
<\infty $. Here, as $\theta \rightarrow \infty $, $\xi _{1}$ is compound
Poisson with rate $1/h^{\prime }\left( 1\right) $.

\emph{Example 2 }Take for $\xi $ the model: $\xi =0$ with probability $1/2$, 
$\xi =2$ with probability $1/2$, so with $\phi \left( z\right) =\left(
1+z^{2}\right) /2.$ Here, provided $n$ is even, 
\begin{equation*}
\mathbf{P}\left( \mathbf{\nu }_{n}=\mathbf{k}_{n}\right) =1/\binom{n}{n/2},%
\text{ }k_{m}\in \left\{ 0,2\right\} \text{, }m=1,...,n\text{, uniform on }%
\left| \mathbf{k}_{n}\right| =n\text{.}
\end{equation*}

\emph{Example 3} In the Moran model, $\mathbf{\nu }_{n}\overset{d}{\sim }$
random permutation of $\left( 2,0,1,...,1\right) :$ in such a model, only
one new gene per generation may come to life, at the expense of the
simultaneous disappearance of some other gene.\newline

In the first two examples, $\mathbf{\nu }_{n}$ is obtained while
conditioning $n$ iid mean $1$ random variables $\mathbf{\xi }_{n}$ on
summing to $n$ (not the case of Moran model) and we shall only consider this
case in the sequel, even assuming (as in the examples) the $\xi _{m}$s to
have finite variance at least: $\sigma ^{2}\left( \xi \right) =\phi ^{\prime
\prime }\left( 1\right) <\infty $. In all such cases, 
\begin{equation*}
\mathbf{P}\left( \mathbf{\nu }_{n}=\mathbf{k}_{n}\right) =\frac{%
\prod_{m=1}^{n}\left[ z^{k_{m}}\right] \phi \left( z\right) }{\left[
z^{n}\right] \phi \left( z\right) ^{n}}\text{, }\left| \mathbf{k}_{n}\right|
=n
\end{equation*}
\begin{equation*}
\mathbf{P}\left( \nu _{1,n}+...+\nu _{k,n}=k^{\prime }\right) =\frac{\left[
z^{k^{\prime }}\right] \phi \left( z\right) ^{k}\left[ z^{n-k^{\prime
}}\right] \phi \left( z\right) ^{n-k}}{\left[ z^{n}\right] \phi \left(
z\right) ^{n}}\text{, }k,k^{\prime }\in \left\{ 0,...,n\right\}
\end{equation*}
\begin{equation*}
\mathbf{E}\left( z^{\nu _{1,n}}\right) =\frac{1}{\left[ z^{n}\right] \phi
\left( z\right) ^{n}}\sum_{k^{\prime }=0}^{n}z^{k^{\prime }}\left[
z^{k^{\prime }}\right] \phi \left( z\right) \left[ z^{n-k^{\prime }}\right]
\phi \left( z\right) ^{n-1}.
\end{equation*}

\subsection{Forward in time branching process (neutral genetic drift)}

Take a sub-sample of size $m$ out of $\left[ n\right] :=\left\{
0,1,...,n\right\} $, at generation $0.$ Given $N_{0}\left( m\right) $ $=m$,
let

\begin{equation*}
N_{r}\left( m\right) \text{ }=\#\text{ offspring at generation }r\in \Bbb{N}%
_{0}\text{, forward-in-time}.
\end{equation*}
$N_{r}\left( m\right) $ represents the descendance at time $r$ of the (say
type $1$) $m$ first individuals of the whole population of size $n$. This
sibship process is a discrete-time homogeneous Markov chain, with transition
probability: 
\begin{equation}
\mathbf{P}\left( N_{r+1}\left( m\right) =k^{\prime }\mid N_{r}\left(
m\right) =k\right) =\mathbf{P}\left( \nu _{1,n}+...+\nu _{k,n}=k^{\prime
}\right) .  \label{Eq1}
\end{equation}
It is a martingale, with state-space $\left\{ 0,...,n\right\} $, initial
state $m$, absorbing states $\left\{ 0,n\right\} $ and transient states $%
\left\{ 1,...,n-1\right\} .$ With $\tau _{m,0}=\inf \left( r:N_{r}\left(
m\right) =0\right) $ and $\tau _{m,n}=\inf \left( r:N_{r}\left( m\right)
=n\right) $, the first hitting time of the boundaries $\left\{ 0,n\right\} $
is: $\tau _{m}=\tau _{m,0}\wedge \tau _{m,n}$. It is finite with probability 
$1$ and has finite mean. Omitting reference to any specific initial
condition $m$, the process $\left( N_{r};r\in \Bbb{N}_{0}\right) $ has the
transition matrix $\Pi _{n}$ with entries $\Pi _{n}\left( k,k^{\prime
}\right) =\mathbf{P}\left( \nu _{1,n}+...+\nu _{k,n}=k^{\prime }\right) $
given by (\ref{Eq1})$.$ We have $\Pi _{n}\left( 0,k^{\prime }\right) =\delta
_{0,k^{\prime }}$ and $\Pi _{n}\left( n,k^{\prime }\right) =\delta
_{n,k^{\prime }}$ and $\Pi _{n}$ is not irreducible. However, $\Pi _{n}$ is
aperiodic and (apart from absorbing states) cannot be broken down into
non-communicating subsets; as a result it is diagonalizable, with
eigenvalues $\lambda _{0}\geq \lambda _{1}\geq \lambda _{2}\geq ...\geq
\lambda _{n}$ and $1=\lambda _{0}=\lambda _{1}>\lambda _{2}$. For $m,k\in
\left\{ 0,n\right\} $, we have 
\begin{equation*}
\mathbf{P}\left( N_{r}\left( m\right) =k\right) =\mathbf{e}_{m}^{\prime }\Pi
_{n}^{r}\mathbf{e}_{k}
\end{equation*}
and therefore, with $\overline{\Pi }_{n}$ the restriction of $\Pi _{n}$ to
the states $\left\{ 1,n-1\right\} $, for $m\in \left\{ 1,n-1\right\} $%
\begin{equation*}
\mathbf{P}\left( \tau _{m}>r\right) =\mathbf{e}_{m}^{\prime }\overline{\Pi }%
_{n}^{r}\mathbf{1,}
\end{equation*}
where $\mathbf{1}$ is the all-one column vector. $\tau _{m}$ has geometric
tails with rate $\lambda _{2}<1$.\newline

\emph{Example 4} (Dirichlet binomial): With $U_{k}$ a $\left( 0,1\right) -$%
valued random variable with density beta$\left( k\theta ,\left( n-k\right)
\theta \right) $%
\begin{equation*}
\mathbf{P}\left( \nu _{1,n}+..+\nu _{k,n}=k^{\prime }\right) =\binom{n}{%
k^{\prime }}\frac{\left[ k\theta \right] _{k^{\prime }}\left[ \left(
n-k\right) \theta \right] _{n-k^{\prime }}}{\left[ n\theta \right] _{n}}=%
\mathbf{E}\left[ \binom{n}{k^{\prime }}U_{k}^{k^{\prime }}\left(
1-U_{k}\right) ^{n-k^{\prime }}\right] ,
\end{equation*}
which is a beta mixture of the binomial distribution Bin$\left( n,u\right) .$
In particular, with $U_{1}$ a $\left( 0,1\right) -$valued random variable
with density beta$\left( \theta ,\left( n-1\right) \theta \right) ,$ we have 
\begin{equation}
\mathbf{E}\left( z^{\nu _{1,n}}\right) =\mathbf{E}\left[ \left(
zU_{1}+1-U_{1}\right) ^{n}\right] .  \label{eq2}
\end{equation}

\emph{Example 5} The Wright-Fisher model has a Bin$\left( n,k/n\right) $
transition matrix: 
\begin{equation*}
\mathbf{P}\left( N_{r+1}\left( m\right) =k^{\prime }\mid N_{r}\left(
m\right) =k\right) =\binom{n}{k^{\prime }}\left( \frac{k}{n}\right)
^{k^{\prime }}\left( 1-\frac{k}{n}\right) ^{n-k^{\prime }}.
\end{equation*}
\textbf{Remarks 1-3}

\textbf{1}/- \textbf{(}statistical symmetry\textbf{)}: Due to
exchangeability of the reproduction law, neutral models are symmetric in the
following sense: The transition probabilities of $\overline{N}_{r}\left(
m\right) :=n-N_{r}\left( m\right) $ are equal to the transition
probabilities of $N_{r}\left( m\right) $. $\overline{N}_{r}\left( m\right) 
\overset{d}{=}N_{r}\left( n-m\right) $ represents the descendance at time $r$
of the (say type $2$) $n-m$ initial individuals complementing $m$ to get the
whole initial population of size $n$.

\textbf{2}/- The smaller $n$, the more the genetic drift process $%
N_{r}\left( m\right) $ looks chaotic as it will reach very fast one of the
absorbing boundary (genetic diversity is reduced fast); the smaller the
population the greater the probability that fluctuations will lead to
absorption. For varying population sizes $n$, to define an effective
(equivalent) population size quantifying the genetic drift, more weight
therefore has to be put on the small values of $n$\ than on the large ones
where the process evolves more smoothly.

\textbf{3}/- As $n\rightarrow \infty $, the finite-dimensional probability
transition matrix $\Pi _{n}\left( k,k^{\prime }\right) =\mathbf{P}\left( \nu
_{1,n}+...+\nu _{k,n}=k^{\prime }\right) $ (with $k,k^{\prime }\in \left[
n\right] $), given by (\ref{Eq1}), tends to the infinite-dimensional
transition matrix $\Pi $ with $\left( k,k^{\prime }\right) $ entries, $%
k,k^{\prime }\in \Bbb{N}_{0}$%
\begin{equation*}
\mathbf{P}\left( N_{r+1}\left( m\right) =k^{\prime }\mid N_{r}\left(
m\right) =k\right) =\Pi \left( k,k^{\prime }\right) =\mathbf{P}\left( \xi
_{1}+...+\xi _{k}=k^{\prime }\right) =\left[ z^{k^{\prime }}\right] \phi
\left( z\right) ^{k}.
\end{equation*}
The latter is the one of a critical branching Galton-Watson process with
offspring distribution $\xi $ given by its pgf $\phi \left( z\right) =%
\mathbf{E}\left( z^{\xi }\right) $. $\Box $

\subsection{Backward in time process (neutral coalescent)}

The coalescent backward count process can be defined as follows: Take a
sub-sample of size $m$ from $\left[ n\right] $ at generation $0.$ Identify
two individuals from $\left[ m\right] $ at each step if they share a common
ancestor one generation backward-in-time. This defines an equivalence
relation between $2$ genes from the set $\left[ m\right] $. With $%
A_{0}\left( m\right) =m$, let

\begin{equation*}
A_{r}\left( m\right) \text{ }=\#\text{ ancestors at generation }r\in \Bbb{N}%
_{0}\text{, backward-in-time}.
\end{equation*}
The backward ancestral count process is a discrete-time Markov chain with
transition probabilities (Cannings, $1974$ and Gladstien, $1978$): 
\begin{equation}
\mathbf{P}\left( A_{r+1}\left( m\right) =a\mid A_{r}\left( m\right)
=b\right) =P_{b,a}^{\left( n\right) }:=\frac{b!}{a!}%
\sum_{b_{1},...,b_{a}}^{*}\frac{P_{b;a}^{\left( n\right) }\left( \mathbf{b}%
_{a}\right) }{b_{1}!...b_{a}!}.  \label{Eq2}
\end{equation}
\begin{equation*}
=\frac{\binom{n}{a}}{\binom{n}{b}}\sum_{b_{1},...,b_{a}}^{*}\mathbf{E}\left(
\prod_{l=1}^{a}\binom{\nu _{l,n}}{b_{l}}\right) .
\end{equation*}
In the latter equations, the star-sum is over $b_{1},...,b_{a}\in \Bbb{N}%
:=\left\{ 1,2,...\right\} $, such that $b_{1}+...+b_{a}=b$. This Markov
chain has state-space $\left\{ 0,...,m\right\} $, initial state $m$,
absorbing states $\left\{ 0,1\right\} .$ The process $\left( A_{r};r\in \Bbb{%
N}_{0}\right) $ has the transition matrix $P_{n}$ with entries $P_{n}\left(
b,a\right) =P_{b,a}^{\left( n\right) }$ given by (\ref{Eq2}). In particular,
the probability that two (three) randomly sampled without replacement
individuals out of $A_{r}\left( m\right) $ share a common ancestor in the
previous generation are, respectively, 
\begin{equation*}
P_{2,1}^{\left( n\right) }=\frac{n\mathbf{E}\left( \binom{\nu _{1,n}}{2}%
\right) }{\binom{n}{2}},\text{ }P_{3,1}^{\left( n\right) }=\frac{n\mathbf{E}%
\left( \binom{\nu _{1,n}}{3}\right) }{\binom{n}{3}}.
\end{equation*}
If $P_{3,1}^{\left( n\right) }/P_{2,1}^{\left( n\right) }\rightarrow 0$ as $%
n\rightarrow \infty $ and $P_{2,1}^{\left( n\right) }=O\left( n^{-1}\right) $%
, the equivalent population size is $n_{e}=1/P_{2,1}^{\left( n\right)
}=O\left( n\right) .$ It can be shown (see e.g. Huillet and M\"{o}hle,
Theorem $2.4$, $2015$) that this situation occurs under the condition $%
\sigma ^{2}\left( \xi \right) =\phi ^{\prime \prime }\left( 1\right) <\infty 
$, and then $n_{e}\sim n/\sigma ^{2}\left( \xi \right) $ for large $n$. In
the P\`{o}lya case for example, $P_{3,1}^{\left( n\right) }/P_{2,1}^{\left(
n\right) }\rightarrow 0$ as $n\rightarrow \infty $ and using (\ref{eq2}), $%
P_{2,1}^{\left( n\right) }=\left( \theta +1\right) /\left( n\theta +1\right) 
$ leading to $n_{e}=\left( n\theta +1\right) /\left( \theta +1\right) $ with 
$n_{e}\sim n/\sigma ^{2}\left( \xi \right) $ and $\sigma ^{2}\left( \xi
\right) =\left( \theta +1\right) /\theta $. Under such conditions, as is
well-known, a space-time scaling limit exists for $N_{r}\left( m\right) $
(the Wright-Fisher diffusion on the unit interval), together with a
time-scaled version of $A_{r}\left( m\right) $ (the Kingman coalescent),
with continuous time measured in units of $n_{e}$. So $n_{e}$ in the context
of the neutral forward dynamics is defined from the backward setup and it
fixes the true time-scale.

\subsection{Duality (neutral case).}

We start with a definition of the duality concept which is relevant to our
purposes.

\textbf{Definition }(Liggett, $1985$): Two Markov processes $\left(
X_{t}^{1},X_{t}^{2};t\geq 0\right) ,$\ with state-spaces $\left( \mathcal{E}%
_{1},\mathcal{E}_{2}\right) ,$\ are said to be dual with respect to some
real-valued function $\Phi $\ on the product space $\mathcal{E}_{1}\times 
\mathcal{E}_{2}$\ if $\forall x_{1}\in \mathcal{E}_{1},$\ $\forall x_{2}\in 
\mathcal{E}_{2},$\ $\forall t\geq 0:$%
\begin{equation}
\mathbf{E}_{x_{1}}\Phi \left( X_{t}^{1},x_{2}\right) =\mathbf{E}_{x_{2}}\Phi
\left( x_{1},X_{t}^{2}\right) .  \label{dual}
\end{equation}
\newline

We then recall basic examples of dual processes from the neutral and
exchangeable population models (M\"{o}hle, $1997$): The neutral forward and
backward processes $\left( N_{r},A_{r};r\in \Bbb{N}_{0}\right) $ introduced
in the two preceding subsections are dual with respect to the hypergeometric
sampling without replacement kernel: 
\begin{equation}
\Phi _{n}\left( m,k\right) =\binom{n-m}{k}/\binom{n}{k}\text{ on }\left\{
0,...,n\right\} ^{2}.  \label{Eq3}
\end{equation}
(\ref{Eq3}) reads: 
\begin{equation*}
\text{ }\mathbf{E}_{m}\left[ \binom{n-N_{r}}{k}/\binom{n}{k}\right] =\mathbf{%
E}_{k}\left[ \binom{n-m}{A_{r}}/\binom{n}{A_{r}}\right] =\mathbf{E}%
_{k}\left[ \binom{n-A_{r}}{m}/\binom{n}{m}\right] .
\end{equation*}
The left-hand-side is the probability that a $k-$sample (without
replacement) from population of size $N_{r}$ at time $r$ are all of type $2$%
, given $N_{0}=m.$ If this $k-$sample are all descendants of $A_{r}$
ancestors at time $-r$, this probability must be equal to the probability
that a $m-$sample from population of size $A_{r}$ at time $-r$ are
themselves all of type $2.$\newline

With $P_{n}^{\prime }$ the transpose of $P_{n}$, a one-step ($r=1$) version
of these formulae is: 
\begin{equation*}
\Pi _{n}\Phi _{n}=\Phi _{n}P_{n}^{\prime }
\end{equation*}
where $\Phi _{n}$ is an $\left( n+1\right) \times \left( n+1\right) $ matrix
with entries $\Phi _{n}\left( m,k\right) $ and $\left( \Pi _{n},P_{n}\right) 
$ the transition matrices of forward and backward processes. Note that the
matrix $\Phi _{n}$ is symmetric and left-upper triangular. The matrix $\Phi
_{n}$ is invertible, with entries 
\begin{equation*}
\Phi _{n}^{-1}\left( i,j\right) =\left( -1\right) ^{i+j-n}\binom{i}{n-j}%
\binom{n}{i}=\left( -1\right) ^{i+j-n}\binom{j}{n-i}\binom{n}{j}.
\end{equation*}
The matrix $\Phi _{n}^{-1}$ is symmetric right-lower triangular. Thus, 
\begin{equation*}
\Phi _{n}^{-1}\Pi _{n}\Phi _{n}=P_{n}^{\prime }.
\end{equation*}
Being similar matrices, $\Pi _{n}$ and $P_{n}^{\prime }$ (or $P_{n}$) both
share the same eigenvalues. In (M\"{o}hle, $1999$), a direct combinatorial
proof of the duality result can be found (in the general exchangeable or
neutral case); it was obtained by directly checking the consistency of (\ref
{Eq1}), (\ref{Eq2}) and (\ref{Eq3}).

The duality formulae allow one to deduce the probabilistic structure of one
process from the one of the other.

\section{Beyond neutrality (symmetry breaking arising from bias)}

Discrete forward non-neutral Markov chain models (with non-null drifts) can
be obtained by substituting 
\begin{equation*}
k\rightarrow np\left( \frac{k}{n}\right) \text{ in }\Pi _{n}\left(
k,k^{\prime }\right) :=\mathbf{P}\left( \nu _{1,n}+...+\nu _{k,n}=k^{\prime
}\right) ,
\end{equation*}
where: 
\begin{equation*}
p\left( x\right) :x\in \left( 0,1\right) \rightarrow \left( 0,1\right) \text{
is continuous, increasing, with }p\left( 0\right) =0,\text{ }p\left(
1\right) =1.
\end{equation*}
$p\left( x\right) $ is the state-dependent Bernoulli bias probability
different from the identity $x$ (as in neutral case). \newline

When particularized to the Wright-Fisher model, this leads to the biased
transition probabilities: 
\begin{equation*}
\mathbf{P}\left( N_{r+1}\left( m\right) =k^{\prime }\mid N_{r}\left(
m\right) =k\right) =\binom{n}{k^{\prime }}p\left( \frac{k}{n}\right)
^{k^{\prime }}\left( 1-p\left( \frac{k}{n}\right) \right) ^{n-k^{\prime }}.
\end{equation*}
In this binomial $n-$sampling with replacement model, a type $1$ individual
is drawn with probability $p\left( \frac{k}{n}\right) $ which is different
from the uniform distribution $k/n$, due to bias effects.

From this, we conclude (a symmetry breaking property): The transition
probabilities of $\overline{N}_{r}\left( m\right) :=n-N_{r}\left( m\right) $%
, $r\in \Bbb{N}_{0}$ are 
\begin{equation*}
\text{Bin}\left( n,1-p\left( 1-k/n\right) \right) \neq \text{Bin}\left(
n,p\left( k/n\right) \right) ,
\end{equation*}
and so, no longer coincide with the ones of $\left( N_{r}\left( m\right)
;r\in \Bbb{N}_{0}\right) .$ The process $N_{r}\left( m\right) $, $r\in \Bbb{N%
}_{0}$, which is Markovian, no longer is a martingale. Rather, if $%
x\rightarrow p\left( x\right) $ is concave (convex), $N_{r}\left( m\right) $%
, $r\in \Bbb{N}_{0}$ is a sub-martingale (super-martingale), because: $%
\mathbf{E}\left( N_{r+1}\left( m\right) \mid N_{r}\left( m\right) \right)
=np\left( N_{r}\left( m\right) /n\right) \geq N_{r}\left( m\right) $
(respectively $\leq N_{r}\left( m\right) $).

In the binomial neutral Wright-Fisher transition probabilities for instance,
we replaced the success probability $k/n$ by a more general function $%
p\left( k/n\right) $. The reproduction law corresponding to the biased
binomial model is multinomial and asymmetric, namely: $\mathbf{\nu }_{n}%
\overset{d}{\sim }$ Multin$\left( n;\mathbf{\pi }_{n}\right) $, where $%
\mathbf{\pi }_{n}:=\left( \pi _{1,n},...,\pi _{n,n}\right) $ and: $\pi
_{m,n}=p\left( m/n\right) -p\left( \left( m-1\right) /n\right) $, $m=1,...,n$
obeying $\sum_{m=1}^{n}\pi _{m,n}=p\left( 1\right) -p\left( 0\right) =1$.
Due to its asymmetry, the law of the biased $\mathbf{\nu }_{n}$ no longer is
exchangeable. We now recall some well-known bias examples arising in
population genetics. \newline

\emph{Example 6} (homographic model, haploid selection). Assume 
\begin{equation}
p\left( x\right) =\left( 1+s\right) x/\left( 1+sx\right) ,  \label{se}
\end{equation}
where $s>-1$ is a selection parameter. This model arises when gene $1$
(respectively $2$), with frequency $x$ (respectively $1-x$), has fitness $%
1+s $ (respectively $1$) in a multiplicative model of fitness. The case $s>0$
arises when gene of type $1$ is selectively advantageous, whereas it is
disadvantageous when $s\in \left( -1,0\right) .$

\emph{Example 7}\textbf{\ }(diploid selection with dominance). Assume 
\begin{equation}
p\left( x\right) =\frac{\left( 1+s\right) x^{2}+\left( 1+sh\right) x\left(
1-x\right) }{1+sx^{2}+2shx\left( 1-x\right) }.  \label{sd}
\end{equation}
In this model, genotype $11$ (respectively $12$ and $22$), with frequency $%
x^{2}$ (respectively $2x\left( 1-x\right) $ and $\left( 1-x\right) ^{2}$)
has fitness $1+s$ (respectively $1+sh$ and $1$). $h$ is a measure of the
degree of dominance of heterozygote $12$. We impose $s>-1$ and $sh>-1.$ Note
that the latter quantity can be put into the canonical form of deviation to
neutrality: 
\begin{equation*}
p\left( x\right) =x+sx\left( 1-x\right) \frac{h-x\left( 2h-1\right) }{%
1+sx^{2}+2shx\left( 1-x\right) },
\end{equation*}
where the ratio appearing in the right-hand-side is the ratio of the
difference of marginal fitnesses of $1$ and $2$ to their mean fitness. The
case $h=1/2$ corresponds to balancing selection with: $p\left( x\right) =x+%
\frac{s}{2}\frac{x\left( 1-x\right) }{1+sx}.$

\emph{Example 8}\textbf{\ }(quadratic segregation model) With $a\in \left[
-1,1\right] ,$ a curvature parameter, one may choose: 
\begin{equation}
p\left( x\right) =x\left( 1+a-ax\right) ,  \label{quad}
\end{equation}
corresponding to a segregation model (Weissing and van Boven, $2001$). If $%
a=1$, $p\left( x\right) =x\left( 2-x\right) =1-\left( 1-x\right) ^{2}$: this
bias appears in a discrete $2$-sex population model (M\"{o}hle, $1994$).

We can relax the assumption $p\left( 0\right) =0,$ $p\left( 1\right) =1$ by
assuming $0\leq p\left( 0\right) \leq $ $p\left( 1\right) \leq 1$, $p\left(
1\right) -p\left( 0\right) \in \left[ 0,1\right) .$

\emph{Example 9}\textbf{\ }(affine mutation model) Take for example 
\begin{equation}
p\left( x\right) =\left( 1-\mu _{2}\right) x+\mu _{1}\left( 1-x\right) ,
\label{mu}
\end{equation}
where $\left( \mu _{1},\mu _{2}\right) $ are mutation probabilities,
satisfying $\mu _{1}\leq 1-\mu _{2}.$ It corresponds to the mutation scheme: 
$1\overset{}{\underset{\mu _{2}}{\overset{\mu _{1}}{\rightleftarrows }}}2$.
To avoid discussions of intermediate cases, we will assume that $p\left(
0\right) =\mu _{1}>0$ and $p\left( 1\right) <1$ ($\mu _{2}>0$)$.$ In this
case, the matrix $\Pi _{n}$ is irreducible and even primitive and all states
of this Markov chain are now recurrent. We have $\mathbf{P}\left(
N_{r+1}>0\mid N_{r}=0\right) =1-\left( 1-p\left( 0\right) \right) ^{n}>0$
and $\mathbf{P}\left( N_{r+1}<n\mid N_{r}=n\right) =1-p\left( 1\right)
^{n}>0 $ and the boundaries $\left\{ 0\right\} $ and $\left\{ n\right\} $ no
longer are strictly absorbing as there is a positive reflection probability
inside the domain $\left\{ 0,1,...,n\right\} $.

\section{A modified Cannings model with constant population size on average}

The condition that to form the next generation the offspring should preserve
exactly the population size to $n$ is a drastic one and we wish here to
discuss a weaker form of conservation, namely conservation of the total
population size on average. And see how the latter theory with constant
population size is modified in depth.\newline

\textbf{The frame (or environment) process.} Consider a population of
initial size $n$ as before. Assume each individual generates a random number
of offspring of size $\xi _{m}$, independently of one another, with the $\xi
_{m}$s iid with $\mathbf{P}\left( \xi =k\right) =:\pi _{k}$. Assume $\mathbf{%
E}\left( \xi \right) =1$ and introduce $\mathbf{E}\left( z^{\xi }\right)
=:\phi \left( z\right) $, the pgf of $\xi $ then with $\phi ^{\prime }\left(
1\right) =1$. We shall moreover assume $\sigma ^{2}\left( \xi \right) =\phi
^{\prime \prime }\left( 1\right) <\infty $ (finiteness of the variance of $%
\xi $). Then we are left with a classical critical branching Galton-Watson
process with finite variance.

Let $N_{r}\left( n\right) $ be the number of descendants at generation $r$,
given initially $N_{0}\left( n\right) =n.$ We have 
\begin{equation}
N_{r+1}\left( n\right) =\sum_{m=1}^{N_{r}\left( n\right) }\xi _{m},
\label{f1}
\end{equation}
so that $\mathbf{E}\left( N_{r+1}\left( n\right) \right) =\mathbf{E}\left(
\xi \right) \mathbf{E}\left( N_{r}\left( n\right) \right) $. Thus for each $%
r\geq 0$, $\mathbf{E}\left( N_{r}\left( n\right) \right) =n$ and the
population size is conserved on average only. We shall call the process $%
N_{r}\left( n\right) $, $r\geq 0$, the frame or environment process. It
models a fluctuating total population size constant on average. Note that $%
n=N_{0}$ could be made random with mean $\mu $ so that in this case $\mathbf{%
E}\left( N_{r}\left( N_{0}\right) \right) =\mathbf{E}\left( N_{0}\right)
=\mu $ but in the sequel, for the sake of simplicity, we shall work
conditionally given $N_{0}=n$.\newline

\textbf{A two types population model.} Let $1\leq m\leq n$ and $N_{r}\left(
m\right) $ be the number of (say type $1$) descendants at generation $r$ of
the $N_{0}\left( m\right) =m$ first founders (as a subset of the full
initial population with $n$ founders). We shall also let $N_{r}\left(
n-m\right) =N_{r}\left( n\right) -N_{r}\left( m\right) $ be the number of
(say type $2$) descendants at generation $r$ of the remaining part of the
initial population. With $\phi _{r}\left( z\right) :=\mathbf{E}\left(
z^{N_{r}\left( 1\right) }\right) $ the pgf of $N_{r}\left( 1\right) $, we
have $\phi _{r+1}\left( z\right) =\phi \left( \phi _{r}\left( z\right)
\right) $, $\phi _{0}\left( z\right) =z$ so that $\phi _{r}\left( z\right)
=\phi ^{\circ r}\left( z\right) $ the $r-$fold composition of $\phi $ with
itself. And by independence of the founders 
\begin{equation}
\left\{ 
\begin{array}{c}
\mathbf{E}\left( z^{N_{r}\left( m\right) }\right) =\phi _{r}\left( z\right)
^{m},\text{ with} \\ 
\mathbf{E}\left( N_{r}\left( m\right) \right) =m\text{ and }\mathbf{E}\left(
N_{r}\left( m\right) ^{2}\right) =m^{2}+rm\phi ^{\prime \prime }\left(
1\right) .
\end{array}
\right.  \label{f2}
\end{equation}
If $r_{2}>r_{1}>0$, using the following expression of the joint pgf of $%
N_{r_{1}}\left( m\right) ,$ $N_{r_{2}}\left( m\right) $, say $\mathbf{E}%
\left( z_{1}^{N_{r_{1}}\left( m\right) }z_{2}^{N_{r_{2}}\left( m\right)
}\right) =\phi _{r_{1}}\left( z_{1}\phi _{r_{2}-r_{1}}\left( z_{2}\right)
\right) ^{m}$, we get 
\begin{eqnarray*}
\text{Cov}\left( N_{r_{1}}\left( m\right) ,N_{r_{2}}\left( m\right) \right)
&=&r_{1}m\phi ^{\prime \prime }\left( 1\right) \text{, independently of }%
r_{2}>r_{1} \\
\text{Corr}\left( N_{r_{1}}\left( m\right) ,N_{r_{2}}\left( m\right) \right)
&=&\frac{\text{Cov}\left( N_{r_{1}}\left( m\right) ,N_{r_{2}}\left( m\right)
\right) }{\sigma \left( N_{r_{1}}\left( m\right) \right) \sigma \left(
N_{r_{1}}\left( m\right) \right) }=\sqrt{\frac{r_{1}}{r_{2}}}=\left( 1+\frac{%
r_{2}-r_{1}}{r_{1}}\right) ^{-1/2},
\end{eqnarray*}
independent of $m$. The process $N_{r}\left( m\right) $ exhibits positive
long-range correlations.

Note that if $m=n$, we are left with the frame process and $m<n\Rightarrow
N_{r}\left( m\right) \leq N_{r}\left( n\right) $, $\forall r$: $N_{r}\left(
m\right) $ is a sub-process of $N_{r}\left( n\right) $. While if $m=1$, the
process $N_{r}\left( 1\right) $, as a special sub-process of $N_{r}\left(
n\right) $, describes the fate of an initial mutant.

The process $N_{r}\left( m\right) $, $r\in \Bbb{N}_{0}$, as a critical
Galton-Watson branching process itself, is a discrete-time Markov chain with
initial condition $N_{0}\left( m\right) =m$, state-space $\Bbb{N}_{0}$, and
(semi-)infinite-dimensional transition matrix 
\begin{equation*}
\mathbf{P}\left( N_{r+1}\left( m\right) =k^{\prime }\mid N_{r}\left(
m\right) =k\right) =\left[ z^{k^{\prime }}\right] \phi \left( z\right)
^{k}=:\Pi ,\text{ }k,k^{\prime }\in \Bbb{N}_{0}.
\end{equation*}
Therefore, with $\mathbf{e}_{m}^{\prime }=\left( 0,...,0,1,0,...,0\right) $
with the $1$ in position $m$, $m=0,1,...$ ($^{\prime }$ denoting
transposition of the column vector $\mathbf{e}_{m}$)$,$%
\begin{equation*}
\mathbf{P}\left( N_{r}\left( m\right) =k\right) =\mathbf{e}_{m}^{\prime }\Pi
^{r}\mathbf{e}_{k}.
\end{equation*}
The process $N_{r}\left( m\right) $ is the process of interest to us, as a
process embedded in the frame process $N_{r}\left( n\right) $ giving the
total population size which is highly fluctuating, although with constant
mean $n$.\newline

\textbf{Type 1 extinction event. }The type $1$ descendants of the $m$
original founders go extinct with probability $1$ because this probability
is $\rho _{r}^{m}$ where $\rho _{r}:=\mathbf{P}\left( N_{r}\left( 1\right)
=0\right) =\phi _{r}\left( 0\right) $ tends (slowly) to $1$ as $r\rightarrow
\infty $ (for a critical Galton-Watson process, extinction is almost sure
and state $0$ is absorbing for $N_{r}\left( m\right) $). Indeed, $\rho _{r}$
obeys 
\begin{equation*}
\rho _{r+1}=\phi \left( \rho _{r}\right) \text{, }\rho _{0}=0.
\end{equation*}
Recalling $\phi \left( 1\right) =\phi ^{\prime }\left( 1\right) =1$ and $%
\phi ^{\prime \prime }\left( 1\right) <\infty $, an order-two Taylor
development of $\phi $ near $z=1$ gives 
\begin{eqnarray*}
\rho _{r+1} &=&1+\phi ^{\prime }\left( 1\right) \left( \rho _{r}-1\right) +%
\frac{1}{2}\phi ^{\prime \prime }\left( 1\right) \left( \rho _{r}-1\right)
^{2} \\
&=&\rho _{r}+\frac{1}{2}\phi ^{\prime \prime }\left( 1\right) \left( \rho
_{r}-1\right) ^{2},
\end{eqnarray*}
leading to (with $a:=\phi ^{\prime \prime }\left( 1\right) /2$) $\rho
_{r}\sim 1-\frac{1}{ar}$ as $r$ is large.

The distribution of the extinction time $\tau _{m,0}:=\inf \left(
r>0:N_{r}\left( m\right) =0\right) $ is therefore given by 
\begin{equation*}
\mathbf{P}\left( \tau _{m,0}\leq r\right) =\mathbf{P}\left( N_{r}\left(
m\right) =0\right) =\left[ z^{0}\right] \left( \phi _{r}\left( z\right)
^{m}\right) =\left( \left[ z^{0}\right] \phi _{r}\left( z\right) \right)
^{m}=\rho _{r}^{m},
\end{equation*}
with Pareto$\left( 1\right) $ heavy tails 
\begin{equation}
\mathbf{P}\left( \tau _{m,0}>r\right) \sim \frac{m}{ar}\text{as }r\text{ is
large}.  \label{f3}
\end{equation}
To summarize, $N_{r}\left( m\right) $ goes slowly to $0$ with probability $1$
(extinction is almost sure) but it takes a long time to do so.\newline

\textbf{Remark 4:} All this is also true of course for the frame process $%
N_{r}\left( n\right) $ itself. As a critical branching process, the process $%
N_{r}\left( n\right) $\ tends to $0$ with probability $1$ as well and it has
a constant mean $\mathbf{E}\left( N_{r}\left( n\right) \right) =n$; it has a
variance that goes to infinity linearly with $r:$%
\begin{equation*}
\sigma ^{2}\left( N_{r}\left( n\right) \right) =rn\phi ^{\prime \prime
}\left( 1\right) .
\end{equation*}
It goes extinct with probability $1$ but it takes a long but finite time $%
\tau _{n,0}$ to do so. We have 
\begin{equation*}
\mathbf{P}\left( \tau _{n,0}>r\right) =1-\phi _{r}\left( 0\right) ^{n}\sim
n/\left( ra\right) ,\text{ for large }r,
\end{equation*}
with persistent heavy tails, non-geometric. In particular, $\mathbf{E}\left(
\tau _{n,0}\right) =\infty $. The pgf of $N_{r}\left( n\right) $ conditioned
on $N_{r}\left( n\right) >0$ being 
\begin{equation*}
\frac{\phi _{r}\left( z\right) ^{n}-\phi _{r}\left( 0\right) ^{n}}{1-\phi
_{r}\left( 0\right) ^{n}},\text{ we get}
\end{equation*}
\begin{equation*}
\mathbf{E}\left( N_{r}\left( n\right) \mid N_{r}\left( n\right) >0\right)
\sim ar,\text{ for large }r
\end{equation*}
with slow algebraic growth of order $r$ (in $r$) and independent of $n$.
Because\ $a=\phi ^{\prime \prime }\left( 1\right) /2<\infty $, it holds as
well that (Harris, $1964$), 
\begin{equation}
\mathbf{P}\left( \frac{N_{r}\left( n\right) }{ar}>x\mid N_{r}\left( n\right)
>0\right) \underset{r\rightarrow \infty }{\rightarrow }e^{-x}\text{, }x>0.
\label{f4}
\end{equation}
Therefore, consistently with the previous statements 
\begin{equation*}
\left\{ 
\begin{array}{c}
\mathbf{E}\left( N_{r}\left( n\right) \right) =0\cdot \mathbf{P}\left( \tau
_{n,0}\leq r\right) +\mathbf{E}\left( N_{r}\left( n\right) \mid N_{r}\left(
n\right) >0\right) \mathbf{P}\left( \tau _{n,0}>r\right) \sim ar\frac{n}{ar}%
=n \\ 
\sigma ^{2}\left( N_{r}\left( n\right) \right) =\sigma ^{2}\left(
N_{r}\left( n\right) \mid N_{r}\left( n\right) >0\right) \mathbf{P}\left(
\tau _{n,0}>r\right) \sim 2\left( ar\right) ^{2}\frac{n}{ar}=nr\phi ^{\prime
\prime }\left( 1\right) .\text{ }\Box
\end{array}
\right.
\end{equation*}
\newline

\textbf{Type 1 fixation event. }The question of fixation of the descendance
of the $m$ founders is more tricky. We could define the fixation time of
type $1$ as $\tau _{m,n}:=\inf \left( r>0:N_{r}\left( m\right) \geq n\right) 
$. But this definition is independent of what type $2$ individuals do.

Rather we shall define the fixation event as follows. Suppose that the
fixation time $\tau _{m,fix}=r$ if and only if generation $r$ is the first
time at which $N_{r}\left( n-m\right) =0$ and $N_{r}\left( m\right) >0$
(when type $2$ goes extinct for the first time while some type $1$
individuals still survive). Then, with $\rho _{r}=\mathbf{P}\left(
N_{r}\left( 1\right) =0\right) $ obeying $\rho _{0}=0$ and $\rho _{r}\sim 1-%
\frac{2}{r\phi ^{\prime \prime }\left( 1\right) }\rightarrow 1$ as $%
r\rightarrow \infty $, 
\begin{equation}
\mathbf{P}\left( \tau _{m,fix}=r\right) =\left( 1-\rho _{r}^{m}\right)
\left( \rho _{r}^{n-m}-\rho _{r-1}^{n-m}\right) \text{, }r\geq 1.  \label{f5}
\end{equation}
The probability that such a fixation event ever occurs therefore is 
\begin{equation}
0<\mathbf{P}\left( \tau _{m,fix}<\infty \right) =\sum_{r\geq 1}\mathbf{P}%
\left( \tau _{m,fix}=r\right) <1\text{.}  \label{f6}
\end{equation}
Note that 
\begin{equation*}
\mathbf{P}\left( \tau _{m,fix}\leq r,N_{r}\left( m\right) >0\right) =\left(
1-\rho _{r}^{m}\right) \rho _{r}^{n-m}\text{, }r\geq 1.
\end{equation*}
Clearly then, with probability $\mathbf{P}\left( \tau _{m,fix}<\infty
\right) $, $\tau _{m,fix}<\tau _{m,0}$ and with complementary probability $%
\mathbf{P}\left( \tau _{m,fix}=\infty \right) =1-\mathbf{P}\left( \tau
_{m,fix}<\infty \right) $, $\tau _{m,0}<\tau _{m,fix}=\infty .$\newline

\textbf{Type }$\mathbf{1}$\textbf{\ population size at fixation.} A natural
question then is: given $\tau _{m,fix}<\infty $ (an event occurring with
probability $\mathbf{P}\left( \tau _{m,fix}<\infty \right) $), what is the
type $1$ population size $N_{\tau _{m,fix}}\left( m\right) $ at the fixation
time event? We have 
\begin{eqnarray*}
\mathbf{E}\left( N_{\tau _{m,fix}}\left( m\right) \right) &=&\sum_{r\geq 1}%
\mathbf{E}\left( N_{r}\left( m\right) \mid N_{r}\left( m\right) >0\right) 
\mathbf{P}\left( \tau _{m,fix}=r\right) \\
&=&m\sum_{r\geq 1}\left( \rho _{r}^{n-m}-\rho _{r-1}^{n-m}\right) =m,
\end{eqnarray*}
the sums telescoping with $\rho _{0}=0$ and $\rho _{\infty }=1$. On average $%
N_{\tau _{m,fix}}\left( m\right) $ lies at $m$. Let us now consider the
variance. We have 
\begin{equation*}
\left[ \mathbf{E}\left( N_{r}\left( m\right) \mid N_{r}\left( m\right)
>0\right) \right] ^{2}=\frac{\left[ \mathbf{E}\left( N_{r}\left( m\right)
\right) \right] ^{2}}{\left( 1-\rho _{r}^{m}\right) ^{2}}=\frac{m^{2}}{%
\left( 1-\rho _{r}^{m}\right) ^{2}}\text{ and}
\end{equation*}

\begin{equation*}
\mathbf{E}\left( N_{r}\left( m\right) ^{2}\mid N_{r}\left( m\right)
>0\right) =\frac{\mathbf{E}\left( N_{r}\left( m\right) ^{2}\right) }{1-\rho
_{r}^{m}}=\frac{m^{2}+rm\phi ^{\prime \prime }\left( 1\right) }{1-\rho
_{r}^{m}}.
\end{equation*}
Therefore, 
\begin{equation*}
\sigma ^{2}\left( N_{\tau _{m,fix}}\left( m\right) \right) =\sum_{r\geq
1}\sigma ^{2}\left( N_{r}\left( m\right) \mid N_{r}\left( m\right) >0\right) 
\mathbf{P}\left( \tau _{m,fix}=r\right)
\end{equation*}
\begin{equation*}
=\sum_{r\geq 1}\left[ \frac{m^{2}+rm\phi ^{\prime \prime }\left( 1\right) }{%
1-\rho _{r}^{m}}-\frac{m^{2}}{\left( 1-\rho _{r}^{m}\right) ^{2}}\right] 
\mathbf{P}\left( \tau _{m,fix}=r\right)
\end{equation*}
\begin{equation*}
=m\phi ^{\prime \prime }\left( 1\right) \sum_{r\geq 1}r\left( \rho
_{r}^{n-m}-\rho _{r-1}^{n-m}\right) -m^{2}\sum_{r\geq 1}\frac{\rho _{r}^{m}}{%
1-\rho _{r}^{m}}\left( \rho _{r}^{n-m}-\rho _{r-1}^{n-m}\right)
\end{equation*}
\begin{equation*}
=m\phi ^{\prime \prime }\left( 1\right) \sum_{r\geq 1}r\left( \rho
_{r}^{n-m}-\rho _{r-1}^{n-m}\right) +m^{2}\mathbf{P}\left( \tau
_{m,fix}=\infty \right) .
\end{equation*}
We obtained 
\begin{equation}
\sigma ^{2}\left( N_{\tau _{m,fix}}\left( m\right) \mid \tau _{m,fix}<\infty
\right) =m\phi ^{\prime \prime }\left( 1\right) \sum_{r\geq 1}r\left( \rho
_{r}^{n-m}-\rho _{r-1}^{n-m}\right) .  \label{f7}
\end{equation}
Let us now show that the sum $\phi ^{\prime \prime }\left( 1\right)
\sum_{r\geq 1}r\left( \rho _{r}^{n-m}-\rho _{r-1}^{n-m}\right) $ is a
divergent one. Recalling the asymptotic shape of $\rho _{r}$, the status of
this sum is given by the status of the integral 
\begin{eqnarray*}
I &=&\int_{1}^{\infty }r\overset{}{\phi ^{\prime \prime }\left( 1\right) 
\frac{d}{dr}\left( \rho _{r}^{n-m}\right) }dr\sim \int_{1}^{\infty }r%
\overset{}{\phi ^{\prime \prime }\left( 1\right) \frac{d}{dr}\left( \left( 1-%
\frac{2}{r\phi ^{\prime \prime }\left( 1\right) }\right) ^{n-m}\right) }dr \\
&=&2\int_{\phi ^{\prime \prime }\left( 1\right) /2}^{\infty }s\overset{}{%
\frac{d}{ds}\left( \left( 1-\frac{1}{s}\right) ^{n-m}\right) }ds=2\left(
n-m\right) \int_{\phi ^{\prime \prime }\left( 1\right) /2}^{\infty }\frac{1}{%
s}\overset{}{\left( 1-\frac{1}{s}\right) ^{n-m-1}}ds \\
&=&2\left( n-m\right) \int_{0}^{2/\phi ^{\prime \prime }\left( 1\right) }%
\frac{1}{u}\overset{}{\left( 1-u\right) ^{n-m-1}}du,
\end{eqnarray*}
which is indeed logarithmically diverging near $0$. Thus $\sigma ^{2}\left(
N_{\tau _{m,fix}}\left( m\right) \right) =\infty $ and $N_{\tau
_{m,fix}}\left( m\right) $ exhibits very large (infinite) fluctuations. The
full pgf of $N_{\tau _{m,fix}}\left( m\right) $ clearly is 
\begin{eqnarray*}
\mathbf{E}\left( z^{N_{\tau _{m,fix}}\left( m\right) }\right) &=&\sum_{r\geq
1}\mathbf{E}\left( z^{N_{r}\left( m\right) }\mid N_{r}\left( m\right)
>0\right) \mathbf{P}\left( \tau _{m,fix}=r\right) \\
&=&\sum_{r\geq 1}\frac{\phi _{r}\left( z\right) ^{m}-\rho _{r}^{m}}{1-\rho
_{r}^{m}}\mathbf{P}\left( \tau _{m,fix}=r\right) \\
&=&\sum_{r\geq 1}\left( \phi _{r}\left( z\right) ^{m}-\rho _{r}^{m}\right)
\left( \rho _{r}^{n-m}-\rho _{r-1}^{n-m}\right) .
\end{eqnarray*}

\textbf{Fixation probability and fixation time distribution. }We can also
estimate the fixation probability for $n\gg m$ large or for $n,m$ large with 
$m/n=\alpha $. From (\ref{f5}), we have 
\begin{equation*}
\mathbf{P}\left( \tau _{m,fix}<\infty \right) =
\end{equation*}
\begin{equation}
\sum_{r\geq 1}\mathbf{P}\left( \tau _{m,fix}=r\right) =\sum_{r\geq 1}\left(
1-\rho _{r}^{m}\right) \left( \rho _{r}^{n-m}-\rho _{r-1}^{n-m}\right)
\label{f7a}
\end{equation}
which can be approximated (observing $\rho _{1}=\phi \left( 0\right) =\pi
_{0}$ and $\rho _{r}\sim 1-\frac{2}{r\phi ^{\prime \prime }\left( 1\right) }$%
) by 
\begin{eqnarray*}
I &=&\int_{1}^{\infty }dr\left( 1-\rho _{r}^{m}\right) \frac{d}{dr}\left(
\rho _{r}^{n-m}\right) =\left( n-m\right) \int_{1}^{\infty }d\rho _{r}\left(
1-\rho _{r}^{m}\right) \rho _{r}^{n-m-1} \\
&=&\left( n-m\right) \int_{\pi _{0}}^{1}d\rho \left( 1-\rho ^{m}\right) \rho
^{n-m-1}=\frac{m}{n}\left( 1-\pi _{0}^{n}\right) -\pi _{0}^{n-m}.
\end{eqnarray*}
of order $m/n$ when $n$ goes large at fixed $m$ or when both $m,n$ go large
at fixed $m/n=\alpha $. This estimate makes sense because in these two
cases, the factor $\rho _{r}^{n-m-1}$ goes to $0$ for small values of $r$ so
that only the large $r$ behavior of $\rho $ contributes where the asymptotic
estimate of $\rho _{r}$ is valid. With $a:=\phi ^{\prime \prime }\left(
1\right) /2$, we also have 
\begin{equation*}
\mathbf{E}\left( \tau _{m,fix}\mid \tau _{m,fix}<\infty \right)
=I^{-1}\sum_{r\geq 1}r\mathbf{P}\left( \tau _{m,fix}=r,\tau _{m,fix}<\infty
\right)
\end{equation*}
\begin{equation}
\sim \frac{\left( n-m\right) }{I}\int_{1}^{\infty }d\rho _{r}r\left( 1-\rho
_{r}^{m}\right) \rho _{r}^{n-m-1}\sim \frac{n-m^{{}}}{aI}\int_{\pi
_{0}}^{1}d\rho \frac{\left( 1-\rho ^{m}\right) \rho ^{n-m-1}}{1-\rho }
\label{f8}
\end{equation}
\begin{equation*}
=\frac{n-m^{{}}}{aI}\sum_{k=1}^{m}\frac{1}{n-m-k-1}\left( 1-\pi
_{0}^{n-m-k-1}\right) <\infty .
\end{equation*}
Note that, proceeding similarly 
\begin{equation}
\mathbf{E}\left( \tau _{m,fix}^{2}\mid \tau _{m,fix}<\infty \right) \sim 
\frac{n-m^{{}}}{aI}\int_{\pi _{0}}^{1}d\rho \frac{1-\rho ^{m}}{\left( 1-\rho
\right) ^{2}}\rho ^{n-m-1},  \label{f9}
\end{equation}
which is a diverging integral near $\rho =1$. The variance of $\tau
_{m,fix}\mid \tau _{m,fix}<\infty $ is infinite.

The formula (\ref{f8}) giving the mean fixation time allows to obtain
asymptotic estimates of this quantity, given fixation occurs. If $n$ goes
large at fixed $m$, we indeed find 
\begin{equation}
\mathbf{E}\left( \tau _{m,fix}\mid \tau _{m,fix}<\infty \right) \sim \frac{%
n^{2}}{ma}\frac{m}{n}=2n/\phi ^{\prime \prime }\left( 1\right) .  \label{f10}
\end{equation}
If both $m,n$ get large at fixed $m/n=\alpha \in \left( 0,1\right) $, we
obtain 
\begin{equation}
\left\{ 
\begin{array}{c}
\mathbf{E}\left( \tau _{m,fix}\mid \tau _{m,fix}<\infty \right) \sim \frac{%
n-m^{{}}}{aI}\int_{0}^{m-1}dk\frac{1-\pi _{0}^{n\left( 1-\rho \right) -k}}{%
n\left( 1-\alpha \right) -k} \\ 
\sim -\frac{2n}{\phi ^{\prime \prime }\left( 1\right) }\frac{1-\alpha }{%
\alpha }\log \left( 1-\alpha \right) .
\end{array}
\right.  \label{f11}
\end{equation}

\textbf{A symmetric definition of }$\tau _{m,ext}$\textbf{\ and }$\tau
_{m,fix}$ \textbf{for type }$\mathbf{1}$\textbf{.} We could have defined
alternatively (and more symmetrically) the extinction time of type $1$ as 
\begin{equation*}
\tau _{m,ext}:=\inf \left( r>0:N_{r}\left( m\right) =0\text{, }N_{r}\left(
n-m\right) >0\right) ,
\end{equation*}
so as the fixation time of type $2$ individuals. And this definition is now
dependent of what type $2$ individuals do. We shall let $\tau _{m}=\tau
_{m,ext}\wedge \tau _{m,fix}$, the global absorption time. To compute the
probability that $\tau _{m,ext}<\infty $ from this new definition of $\tau
_{m,ext}$, it suffices to substitute $n-m$\ to $m$\ in the expression of $%
\mathbf{P}\left( \tau _{m,fix}<\infty \right) $\ for type $1$\ individuals
as computed in (\ref{f7a}).

And clearly, $\tau _{m,ext}\vee \tau _{m,fix}=\infty $, $\tau
_{m,ext}<\infty \Rightarrow \tau _{m,ext}<\tau _{m,fix}=\infty $\ and $\tau
_{m,fix}<\infty \Rightarrow \tau _{m,fix}<\tau _{m,ext}=\infty $.

We have $\mathbf{P}\left( \tau _{m,fix}<\infty \right) +\mathbf{P}\left(
\tau _{m,ext}<\infty \right) <1$, and the missing mass is the probability
that we have neither a fixation nor an extinction event of type $1$, which
happens whenever both type $1$ and $2$ populations die out simultaneously.
This (rare) event occurs with probability 
\begin{equation*}
\theta :=\sum_{r\geq 1}\left( \rho _{r}^{m}-\rho _{r-1}^{m}\right) \left(
\rho _{r}^{n-m}-\rho _{r-1}^{n-m}\right) ,
\end{equation*}
and it is a case when both $\tau _{m,ext}$ and $\tau _{m,fix}=\infty $
(neither fixation nor extinction are achieved and $\theta =\mathbf{P}\left(
\tau _{m,ext}=\tau _{m,fix}=\infty \right) $). Yet, given this last event
has not occurred, we now know how to compute the probability that a fixation
event of type $1$ precedes an extinction event and conversely, normalizing
the unconditional fixation/extinction probabilities by $1-\theta $.

Define next 
\begin{equation*}
\mathbf{P}\left( k_{1}^{\prime },k^{\prime }\mid k_{1},k\right) :=\mathbf{P}%
\left( N_{r+1}\left( m\right) =k_{1}^{\prime },N_{r+1}\left( n\right)
=k^{\prime }\mid N_{r}\left( m\right) =k_{1},N_{r}\left( n\right) =k\right) =
\end{equation*}

\begin{equation*}
\mathbf{P}\left( \xi _{1}+..+\xi _{k_{1}}=k_{1}^{\prime }\right) \mathbf{P}%
\left( \xi _{k_{1}+1}+..+\xi _{k}=k^{\prime }-k_{1}^{\prime }\right) =\left[
z^{k_{1}^{\prime }}\right] \phi \left( z\right) ^{k_{1}}\left[ z^{k^{\prime
}-k_{1}^{\prime }}\right] \phi \left( z\right) ^{k-k_{1}}.
\end{equation*}
Note that $k_{1}=k\Rightarrow \mathbf{P}\left( k_{1}^{\prime },k^{\prime
}\mid k_{1},k\right) =\left[ z^{k^{\prime }}\right] \phi \left( z\right)
^{k}\cdot \delta _{k^{\prime }}\left( k_{1}^{\prime }\right) $ (type $1$ is
fixed), whereas $k_{1}=0\Rightarrow \mathbf{P}\left( k_{1}^{\prime
},k^{\prime }\mid k_{1},k\right) =\left[ z^{k^{\prime }}\right] \phi \left(
z\right) ^{k}\cdot \delta _{0}\left( k_{1}^{\prime }\right) $ (type $2$ is
fixed). 

We have 
\begin{equation*}
\mathbf{P}\left( N_{r+1}\left( m\right) =k_{1}^{\prime }\mid N_{r}\left(
m\right) =k_{1},N_{r}\left( n\right) =k\right) =\sum_{k^{\prime }\geq
k_{1}^{\prime }}\mathbf{P}\left( k_{1}^{\prime },k^{\prime }\mid
k_{1},k\right) 
\end{equation*}
\begin{equation*}
=:\mathbf{P}\left( k_{1}^{\prime }\mid k_{1},k\right) =\left( \left[
z^{k_{1}^{\prime }}\right] \phi \left( z\right) ^{k_{1}}\right) \phi \left(
1\right) ^{k-k_{1}}=\left[ z^{k_{1}^{\prime }}\right] \phi \left( z\right)
^{k_{1}},
\end{equation*}
independent of $k$. Thus $\mathbf{E}\left( z^{N_{r+1}\left( m\right) }\mid
k_{1},k\right) =\phi \left( z\right) ^{k_{1}}$, independent of $k$. In
particular, 
\begin{eqnarray*}
\mathbf{E}\left( N_{r+1}\left( m\right) \mid k_{1},k\right) 
&=&\sum_{k_{1}^{\prime }}k_{1}^{\prime }\left[ z^{k_{1}^{\prime }}\right]
\phi \left( z\right) ^{k_{1}}=\phi \left( 1\right) k_{1}\phi ^{\prime
}\left( 1\right) =k_{1}=k\frac{k_{1}}{k} \\
\sigma ^{2}\left( N_{r+1}\left( m\right) \mid k_{1},k\right)  &=&k_{1}\phi
^{\prime \prime }\left( 1\right) =k\frac{k_{1}}{k}\phi ^{\prime \prime
}\left( 1\right) ,
\end{eqnarray*}
and $N_{r}\left( m\right) $ is a martingale. Similarly, using the
convolution formula, 
\begin{equation*}
\mathbf{P}\left( N_{r+1}\left( n\right) =k^{\prime }\mid N_{r}\left(
m\right) =k_{1},N_{r}\left( n\right) =k\right) =\sum_{k_{1}^{\prime }\leq
k^{\prime }}\mathbf{P}\left( k_{1}^{\prime },k^{\prime }\mid k_{1},k\right) 
\end{equation*}
\begin{equation*}
=:\mathbf{P}\left( k^{\prime }\mid k_{1},k\right) =\sum_{0\leq k_{1}^{\prime
}\leq k^{\prime }}\left[ z^{k_{1}^{\prime }}\right] \phi \left( z\right)
^{k_{1}}\left[ z^{k^{\prime }-k_{1}^{\prime }}\right] \phi \left( z\right)
^{k-k_{1}}=\left[ z^{k^{\prime }}\right] \phi \left( z\right) ^{k},
\end{equation*}
independent of $k_{1}$. This illustrates that both $N_{r}\left( m\right) $
and $N_{r}\left( n\right) $, although not independent, are both Markov
processes.

We also have consistently 
\begin{equation*}
\mathbf{E}\left( z_{1}^{N_{r+1}\left( m\right) }z^{N_{r+1}\left( n\right)
}\mid k_{1},k\right) =\sum_{k_{1}^{\prime }\geq 0}z_{1}^{k_{1}^{\prime
}}\left[ z_{1}^{k_{1}^{\prime }}\right] \phi \left( z_{1}\right)
^{k_{1}}\sum_{k^{\prime }\geq k_{1}^{\prime }}z^{k^{\prime }}\left[
z^{k^{\prime }-k_{1}^{\prime }}\right] \phi \left( z\right) ^{k-k_{1}}
\end{equation*}
\begin{equation}
=\sum_{k_{1}^{\prime }\geq 0}\left( zz_{1}\right) ^{k_{1}^{\prime }}\left[
z_{1}^{k_{1}^{\prime }}\right] \phi \left( z_{1}\right) ^{k_{1}}\phi \left(
z\right) ^{k-k_{1}}=\phi \left( zz_{1}\right) ^{k_{1}}\phi \left( z\right)
^{k-k_{1}},  \label{f12}
\end{equation}
translating the fact that $N_{r}\left( m\right) $, $N_{r}\left( n\right) $
are jointly Markov as well (with Markov marginals).\newline

\textbf{Introducing bias (deviation to neutrality):} We now wish to
introduce bias translating some interaction between type $1$ and type $2$
populations and see what changes as compared to the neutral case. For some
bias function $p:$ $\left[ 0,1\right] \rightarrow \left[ 0,1\right] $,
different from the identity, such as the ones introduced in Section $3$, let 
\begin{equation}
\begin{array}{c}
\mathbf{P}\left( N_{r+1}\left( m\right) =k_{1}^{\prime },N_{r+1}\left(
n\right) =k^{\prime }\mid N_{r}\left( m\right) =k_{1},N_{r}\left( n\right)
=k\right) \\ 
=\left[ z^{k_{1}^{\prime }}\right] \phi \left( z\right) ^{kp\left(
k_{1}/k\right) }\left[ z^{k^{\prime }-k_{1}^{\prime }}\right] \phi \left(
z\right) ^{k\left( 1-p\left( k_{1}/k\right) \right) },
\end{array}
\label{f13}
\end{equation}
be the transition probability matrix of the joint process $N_{r}\left(
m\right) $, $N_{r}\left( n\right) $, now with bias $p$. Then 
\begin{equation*}
\mathbf{P}\left( N_{r+1}\left( m\right) =k_{1}^{\prime }\mid N_{r}\left(
m\right) =k_{1},N_{r}\left( n\right) =k\right) =\sum_{k^{\prime }\geq
k_{1}^{\prime }}\mathbf{P}\left( k_{1}^{\prime },k^{\prime }\mid
k_{1},k\right)
\end{equation*}
\begin{equation*}
=:\mathbf{P}\left( k_{1}^{\prime }\mid k_{1},k\right) =\left( \left[
z^{k_{1}^{\prime }}\right] \phi \left( z\right) ^{kp\left( k_{1}/k\right)
}\right) \phi \left( 1\right) ^{k\left( 1-p\left( k_{1}/k\right) \right)
}=\left[ z^{k_{1}^{\prime }}\right] \phi \left( z\right) ^{kp\left(
k_{1}/k\right) },
\end{equation*}
Thus $\mathbf{E}\left( z_{1}^{N_{r+1}\left( m\right) }\mid k_{1},k\right)
=\phi \left( z_{1}\right) ^{kp\left( k_{1}/k\right) }$, now dependent on
both $k_{1},k$. In particular, 
\begin{equation}
\left\{ 
\begin{array}{c}
\mathbf{E}\left( N_{r+1}\left( m\right) \mid k_{1},k\right) =kp\left(
k_{1}/k\right) \\ 
\sigma ^{2}\left( N_{r+1}\left( m\right) \mid k_{1},k\right) =kp\left(
k_{1}/k\right) \phi ^{\prime \prime }\left( 1\right) .
\end{array}
\right.  \label{f14}
\end{equation}
The marginal process $N_{r}\left( m\right) $ is no longer Markov, nor is it
a martingale anymore as a result of introducing bias. We also have,
consistently, 
\begin{equation*}
\mathbf{E}\left( z_{1}^{N_{r+1}\left( m\right) }z^{N_{r+1}\left( n\right)
}\mid k_{1},k\right) =
\end{equation*}
\begin{equation*}
=\sum_{k_{1}^{\prime }\geq 0}z_{1}^{k_{1}^{\prime }}\left[
z_{1}^{k_{1}^{\prime }}\right] \phi \left( z_{1}\right) ^{kp\left(
k_{1}/k\right) }\sum_{k^{\prime }\geq k_{1}^{\prime }}z^{k^{\prime }}\left[
z^{k^{\prime }-k_{1}^{\prime }}\right] \phi \left( z\right) ^{k\left(
1-p\left( k_{1}/k\right) \right) }=
\end{equation*}
\begin{equation*}
\sum_{k_{1}^{\prime }\geq 0}\left( zz_{1}\right) ^{k_{1}^{\prime }}\left[
z_{1}^{k_{1}^{\prime }}\right] \phi \left( z_{1}\right) ^{kp\left(
k_{1}/k\right) }\phi \left( z\right) ^{k\left( 1-p\left( k_{1}/k\right)
\right) }=\phi \left( zz_{1}\right) ^{kp\left( k_{1}/k\right) }\phi \left(
z\right) ^{k\left( 1-p\left( k_{1}/k\right) \right) },
\end{equation*}
translating that $N_{r}\left( m\right) $, $N_{r}\left( n\right) $ are still
jointly Markov though. Note $\mathbf{E}\left( z^{N_{r+1}\left( n\right)
}\mid k_{1},k\right) =\phi \left( z\right) ^{k}$, independent of $k_{1}$:
the frame process $N_{r}\left( n\right) $ still is a Markov critical
Galton-Watson process. And, with $N_{r}\left( n-m\right) =N_{r}\left(
n\right) -N_{r}\left( m\right) $, given $N_{r}\left( n-m\right)
=k_{2}=k-k_{1}$ and $N_{r}\left( n\right) =k$, $\mathbf{E}\left(
z_{2}^{N_{r+1}\left( n-m\right) }\mid k_{2},k\right) =\phi \left(
z_{2}\right) ^{k\left( 1-p\left( 1-k_{2}/k\right) \right) }$, dependent on
both $k_{2},k$. Type $2$ $N_{r}\left( n-m\right) $ is non-Markov either.%
\newline

\textbf{Effective population size:} when $n$ is a fixed population size, $%
c_{n}=O\left( 1/n\right) $ (as in the P\`{o}lya model for example) is the
probability that two distinct randomly chosen individuals out of $n$ have
the same direct ancestor. If the population profile is $N_{r}\left( n\right) 
$ and so variable, $c_{n}$ should be replaced by $\overline{c}%
_{n}=R^{-1}\sum_{r=1}^{R}1/N_{r}\left( n\right) $, the empirical average of
the $c_{N_{r}\left( n\right) }$s over some (maybe long) time period $R$. The
effective population size $n_{e}=1/c_{n}$ should accordingly be replaced by 
\begin{equation*}
n_{e}=1/\overline{c}_{n}=\frac{R}{\sum_{r=1}^{R}1/N_{r}\left( n\right) },
\end{equation*}
the harmonic mean of the $N_{r}\left( n\right) $s. The effective population
size used when there are changes in population sizes is thus the harmonic
mean (rather than the arithmetic one). This reflects that if a population
has recovered from a bottleneck (a dramatic reduction in population size), a
definition of the effective population size after a bottleneck should
enhance instants where $N_{r}\left( n\right) $ is small. Because genetic
drift acts more quickly to reduce genetic variation in small populations
(see Remark $2$), genetic diversity in a population will indeed be
substantially reduced when the population size shrinks in a bottleneck event
(the founder effect).

Our construction with variable population size is conditionally given $%
N_{0}\left( n\right) =n$, where $N_{r}\left( n\right) $ is a critical
Galton-Watson sequence of random variables giving the random size of the
global population in each generation $r$. Averaging over the $N_{r}\left(
n\right) $s, while taking into account that $N_{r}\left( n\right) $ goes
extinct with probability $1$ at time $\tau _{n,0}=\inf \left( r\geq
1:N_{r}\left( n\right) =0\right) $, we can define the effective population
size $n_{e}$ in our context by 
\begin{equation}
n_{e}=\mathbf{E}\frac{\tau _{n,0}}{\sum_{r=0}^{\tau _{n,0}-1}1/N_{r}\left(
n\right) }<\mathbf{E}\frac{\sum_{r=0}^{\tau _{n,0}-1}N_{r}\left( n\right) }{%
\tau _{n,0}}=an\mathbf{P}\left( \tau _{n,0}<\infty \right) =an.  \label{f15}
\end{equation}
We used here that the harmonic mean is dominated by the arithmetic one. 
\newline

\textbf{Survival probability to a bottleneck effect. }A bottleneck effect
occurs when the total population size shrinks to a small value before
recovering. We wish to compute the survival probability of type $1$
individuals in such a situation. More precisely, let $k_{1},k_{2}$ obeying $%
k_{1}<<k_{2}$ be the (bottom and top) total population sizes at $%
r_{1}<r_{2}, $ respectively. We may assume $k_{2}$ of order $n$, the mean
value of the total population size at $r_{2},$ whereas $k_{1}$ is assumed
comparatively small because of shrinkage at $r_{1}$. We therefore wish to
compute 
\begin{equation*}
\mathbf{P}\left( N_{r_{2}}\left( m\right) >0\mid N_{r_{1}}\left( n\right)
=k_{1},N_{r_{2}}\left( n\right) =k_{2}\right) .
\end{equation*}
We have 
\begin{equation*}
\mathbf{E}\left( z_{1}^{N_{r_{1}}\left( m\right) }z_{2}^{N_{r_{2}}\left(
m\right) }z_{3}^{N_{r_{1}}\left( n-m\right) }z_{4}^{N_{r_{2}}\left(
n-m\right) }\right) =\phi _{r_{1}}\left( z_{1}\phi _{r_{2}-r_{1}}\left(
z_{2}\right) \right) ^{m}\phi _{r_{1}}\left( z_{3}\phi _{r_{2}-r_{1}}\left(
z_{4}\right) \right) ^{n-m},
\end{equation*}
therefore (substituting $z_{2}z_{4}$ to $z_{2}$ and taking $z_{1}=z_{3}$) 
\begin{equation}
\begin{array}{c}
\Phi _{r_{1},r_{2}}\left( z_{2},z_{3},z_{4}\right) :=\mathbf{E}\left(
z_{2}^{N_{r_{2}}\left( m\right) }z_{3}^{N_{r_{1}}\left( n\right)
}z_{4}^{N_{r_{2}}\left( n\right) }\right) \\ 
=\phi _{r_{1}}\left( z_{3}\phi _{r_{2}-r_{1}}\left( z_{2}z_{4}\right)
\right) ^{m}\phi _{r_{1}}\left( z_{3}\phi _{r_{2}-r_{1}}\left( z_{4}\right)
\right) ^{n-m}.
\end{array}
\label{f16}
\end{equation}
We thus get 
\begin{equation*}
\mathbf{P}\left( N_{r_{2}}\left( m\right) >0\mid N_{r_{1}}\left( n\right)
=k_{1},N_{r_{2}}\left( n\right) =k_{2}\right)
\end{equation*}
\begin{equation*}
=1-\frac{\left[ z_{2}^{0}z_{3}^{k_{1}}z_{4}^{k_{2}}\right] \Phi
_{r_{1},r_{2}}\left( z_{2},z_{3},z_{4}\right) }{\left[
z_{3}^{k_{1}}z_{4}^{k_{2}}\right] \Phi _{r_{1},r_{2}}\left(
1,z_{3},z_{4}\right) }=1-\frac{\left[ z_{3}^{k_{1}}z_{4}^{k_{2}}\right] \Phi
_{r_{1},r_{2}}\left( 0,z_{3},z_{4}\right) }{\left[
z_{3}^{k_{1}}z_{4}^{k_{2}}\right] \Phi _{r_{1},r_{2}}\left(
1,z_{3},z_{4}\right) },
\end{equation*}
where 
\begin{eqnarray*}
\Phi _{r_{1},r_{2}}\left( 1,z_{3},z_{4}\right) &=&\phi _{r_{1}}\left(
z_{3}\phi _{r_{2}-r_{1}}\left( z_{4}\right) \right) ^{n} \\
\Phi _{r_{1},r_{2}}\left( 0,z_{3},z_{4}\right) &=&\phi _{r_{1}}\left(
z_{3}\phi _{r_{2}-r_{1}}\left( 0\right) \right) ^{m}\phi _{r_{1}}\left(
z_{3}\phi _{r_{2}-r_{1}}\left( z_{4}\right) \right) ^{n-m}.
\end{eqnarray*}
We have 
\begin{eqnarray*}
\left[ z_{3}^{k_{1}}\right] \Phi _{r_{1},r_{2}}\left( 1,z_{3},z_{4}\right)
&=&\phi _{r_{2}-r_{1}}\left( z_{4}\right) ^{k_{1}}\cdot \left[
z_{3}^{k_{1}}\right] \phi _{r_{1}}\left( z_{3}\right) ^{n} \\
\left[ z_{3}^{k_{1}}z_{4}^{k_{2}}\right] \Phi _{r_{1},r_{2}}\left(
1,z_{3},z_{4}\right) &=&\left[ z_{3}^{k_{1}}\right] \phi _{r_{1}}\left(
z_{3}\right) ^{n}\cdot \left[ z_{4}^{k_{2}}\right] \phi _{r_{2}-r_{1}}\left(
z_{4}\right) ^{k_{1}}
\end{eqnarray*}
\begin{equation*}
\left[ z_{3}^{k_{1}}\right] \Phi _{r_{1},r_{2}}\left( 0,z_{3},z_{4}\right)
=\sum_{k=0}^{k_{1}}\left[ z_{3}^{k}\right] \phi _{r_{1}}\left( z_{3}\phi
_{r_{2}-r_{1}}\left( z_{4}\right) \right) ^{n-m}\cdot \left[
z_{3}^{k_{1}-k}\right] \phi _{r_{1}}\left( z_{3}\phi _{r_{2}-r_{1}}\left(
0\right) \right) ^{m}
\end{equation*}
\begin{equation*}
=\sum_{k=0}^{k_{1}}\phi _{r_{2}-r_{1}}\left( z_{4}\right) ^{k}\left[
z_{3}^{k}\right] \phi _{r_{1}}\left( z_{3}\right) ^{n-m}\phi
_{r_{2}-r_{1}}\left( 0\right) ^{k_{1}-k}\left[ z_{3}^{k_{1}-k}\right] \phi
_{r_{1}}\left( z_{3}\right) ^{m}
\end{equation*}
\begin{equation*}
\left[ z_{3}^{k_{1}}z_{4}^{k_{2}}\right] \Phi _{r_{1},r_{2}}\left(
0,z_{3},z_{4}\right) =
\end{equation*}
\begin{equation*}
\sum_{k=0}^{k_{1}}\left[ z_{4}^{k_{2}}\right] \phi _{r_{2}-r_{1}}\left(
z_{4}\right) ^{k}\cdot \left[ z_{3}^{k}\right] \phi _{r_{1}}\left(
z_{3}\right) ^{n-m}\phi _{r_{2}-r_{1}}\left( 0\right) ^{k_{1}-k}\cdot \left[
z_{3}^{k_{1}-k}\right] \phi _{r_{1}}\left( z_{3}\right) ^{m}.
\end{equation*}
We thus only need to compute $\left[ z^{k}\right] \phi _{r}\left( z\right)
^{n}$ for various values of $k,n$\footnote{%
Following (Gardy, $1995$), useful asymptotic estimates of such large powers
quantities under different regimes for $k,n$ are available, namely: $k$
fixed, $n\rightarrow \infty ,$ $k=o\left( n\right) $ and $n\rightarrow
\infty $ or $k=O\left( n\right) $ and $n\rightarrow \infty $.}. With $%
f\left( z\right) =\left( \phi _{r}\left( 0\right) +z\right) ^{n}$, $g\left(
z\right) =\phi _{r}\left( z\right) -\phi _{r}\left( 0\right) $, we have 
\begin{equation*}
\phi _{r}\left( z\right) ^{n}=f\left( g\left( z\right) \right) .
\end{equation*}
By Fa\`{a} di Bruno formula (see e.g. Comtet, $1970$, Tome $1$, p. $148$),
with $B_{k,l}\left( g_{1},g_{2},...\right) $ the Bell polynomials in the
Taylor coefficients $g_{k}$ of $g\left( z\right) $: $g_{k}=k!\left[
z^{k}\right] g\left( z\right) =k!\mathbf{P}\left( N_{r}\left( 1\right)
=k\right) $, 
\begin{equation*}
\left[ z^{k}\right] \phi _{r}\left( z\right) ^{n}=:a_{k,n}\left( r\right) =%
\frac{1}{k!}\sum_{l=1}^{k}\left( n\right) _{l}\phi _{r}\left( 0\right)
^{n-l}B_{k,l}\left( g_{1},g_{2},...\right) .
\end{equation*}
So we end up with a combinatorial closed form formula for the survival
probability to a bottleneck effect as\footnote{%
Observing: $\phi _{r}\left( 0\right) :=\left[ z^{0}\right] \phi _{r}\left(
z\right) =a_{0,1}\left( r\right) .$}: 
\begin{equation}
\begin{array}{c}
\mathbf{P}\left( N_{r_{2}}\left( m\right) >0\mid N_{r_{1}}\left( n\right)
=k_{1},N_{r_{2}}\left( n\right) =k_{2}\right) = \\ 
1-\frac{\sum_{k=0}^{k_{1}}a_{k_{2},k}\left( r_{2}-r_{1}\right)
a_{k,n-m}\left( r_{1}\right) a_{k_{1}-k,m}\left( r_{1}\right) a_{0,1}\left(
r_{2}-r_{1}\right) ^{k_{1}-k}}{a_{k_{1},n}\left( r_{1}\right)
a_{k_{2},k_{1}}\left( r_{2}-r_{1}\right) }.
\end{array}
\label{f17}
\end{equation}
\newline

$\emph{Example}$ $\emph{10}$ With $p_{0}+q_{0}=p+q=1$, suppose the critical
homographic model $\phi \left( z\right) =q_{0}+p_{0}\frac{qz}{1-pz}$, a
mixture of a Bernoulli$\left( p_{0}\right) $ random variable with a geometric%
$\left( p\right) $ one. We have $\phi ^{\prime }\left( 1\right) =1$ if $%
p_{0}=q$ and then $\phi ^{^{\prime \prime }}\left( 1\right) =2p/q=:2a$. Thus

\begin{eqnarray*}
\phi \left( z\right) &=&1-\left( \left( 1-z\right) ^{-1}+a\right) ^{-1}\text{
and } \\
\phi _{r}\left( z\right) &=&1-\left( \left( 1-z\right) ^{-1}+ra\right) ^{-1}%
\text{ or }\phi _{r}\left( z\right) =\frac{ra+z\left( 1-ra\right) }{1+ra-raz}%
.
\end{eqnarray*}
Here, $\left[ z^{0}\right] \phi _{r}\left( z\right) ^{n}=\left( \frac{ra}{%
1+ra}\right) ^{n}$ and if $k\geq 1,$%
\begin{equation}
\left\{ 
\begin{array}{c}
\left[ z^{k}\right] \phi _{r}\left( z\right) ^{n}=:a_{k,n}\left( r\right) \\ 
=\sum_{l=1}^{n}\binom{n}{l}\binom{k+l-1}{k}\left( \frac{ra}{1+ra}\right)
^{k}\left( \frac{ra-1}{ra}\right) ^{n-l}\left( ra\left( 1+ra\right) \right)
^{-l} \\ 
=\left( \frac{ra}{ra+1}\right) ^{k}\left( \frac{ra-1}{ra}\right)
^{n}\sum_{l=1}^{n}\binom{n}{l}\binom{k+l-1}{k}\left( \left( ra\right)
^{2}-1\right) ^{-l},
\end{array}
\right.  \label{f18}
\end{equation}
which can be expressed in terms of an hypergeometric function $_{2}F_{1}$.

This homographic (or linear-fractional) model is of particular interest in
our context for two reasons:

1/ One is because $\phi _{r}\left( z\right) $, as the $r$-th composition of $%
\phi \left( z\right) $ with itself, is explicitly computable, as is
well-known from the theory of branching processes (the invariance under
iterated composition property; Harris, $1963$, p. $9$). $\phi \left(
z\right) $ is the pgf of an infinitely divisible random variable $\xi $ if
and only if $p\geq \sqrt{2}-1$ or $\sigma ^{2}\left( \xi \right) =\phi
^{\prime \prime }\left( 1\right) \geq \sqrt{2}$ (Steutel and van Harn,
Example $11.15$, $2003$).

2/ Because all critical branching processes generated by some random
variable $\xi $ are in the domain of attraction of the critical homographic
one. Indeed, with $\phi \left( z\right) $ the pgf of $\xi $, the pgf $\phi
_{r}\left( z\right) $ of $N_{r}\left( 1\right) $ obeys 
\begin{equation*}
\phi _{r+1}\left( z\right) =\phi \left( \phi _{r}\left( z\right) \right) 
\text{, }\phi _{0}\left( z\right) =z.
\end{equation*}
When $r$ gets large, due to almost sure extinction, $\phi _{r}\left(
z\right) $ approaches $1$ (the pgf of $N_{r}\left( 1\right) =0$). Recalling $%
\phi \left( 1\right) =\phi ^{\prime }\left( 1\right) =1$ and $\phi ^{\prime
\prime }\left( 1\right) <\infty $, an order-two Taylor development of $\phi $
near $z=1$ therefore gives 
\begin{eqnarray*}
\phi _{r+1}\left( z\right) &=&1+\phi ^{\prime }\left( 1\right) \left( \phi
_{r}\left( z\right) -1\right) +\frac{1}{2}\phi ^{\prime \prime }\left(
1\right) \left( \phi _{r}\left( z\right) -1\right) ^{2} \\
&=&\phi _{r}\left( z\right) +\frac{1}{2}\phi ^{\prime \prime }\left(
1\right) \left( \phi _{r}\left( z\right) -1\right) ^{2},
\end{eqnarray*}
leading (recalling $a:=\phi ^{\prime \prime }\left( 1\right) /2$) to 
\begin{equation*}
\phi _{r}\left( z\right) \sim 1-1/\left( \left( 1-z\right) ^{-1}+ra\right) ,%
\text{ as }r\text{ is large. }\Box
\end{equation*}
\textbf{Acknowledgments:}

T. Huillet acknowledges partial support from the ``Chaire \textit{%
Mod\'{e}lisation math\'{e}matique et biodiversit\'{e}''.} N. Grosjean and T.
Huillet also acknowledge support from the labex MME-DII Center of Excellence
(\textit{Mod\`{e}les math\'{e}matiques et \'{e}conomiques de la dynamique,
de l'incertitude et des interactions}, ANR-11-LABX-0023-01 project).

\end{document}